\documentclass[12pt]{article}
\usepackage[utf8]{inputenc}
\usepackage[margin=1in]{geometry}
\usepackage{setspace}
\usepackage{color}
\usepackage{titling}
\usepackage{graphicx}
\usepackage[bf,margin=20pt,font={small}]{caption}
\usepackage{tikz}
\usetikzlibrary{math}
\usepackage{amsmath,amssymb,dsfont}
\makeatletter\let\over\@@over\makeatother    %To suppress the \over warning
\makeatletter\let\atop\@@atop\makeatother
\usepackage{hyperref}

\newcommand\TL{\hfil$\displaystyle{##}$}
\newcommand\TR{$\displaystyle{{}##}$\hfil}

\newcommand\TT{\hbox{##}}
\def\seqalign#1#2{\vcenter{\openup1\jot
  \halign{\strut #1\cr #2 \cr}}}
\def\lbldef#1#2{\expandafter\gdef\csname #1\endcsname {#2}}
\newcommand{\eqn}[3][]{\lbldef{#2}{(\ref{#2})}%
\begin{equation} \eqalign{#3} \label{#2} \end{equation}}
\def\eqalign#1{\vcenter{\openup1\jot
    \halign{\strut\span\TL & \span\TR\cr #1 \cr
   }}}
\def\eno#1{(\ref{#1})}
\def\mop#1{\mathop{\rm #1}\nolimits}
\def\sgn{\mop{sgn}}
\def\mod{\mop{\rm mod}}  %Because default definition gives too much space on the left
\def\tr{\mop{tr}}
\def\Im{\mop{Im}}

\renewenvironment{abstract}
 {\normalsize
  \begin{center}
   \bfseries \abstractname\vspace{-.5em}\vspace{0pt}
  \end{center}
  \list{}{
   \setlength{\leftmargin}{0in}%
   \setlength{\rightmargin}{\leftmargin}%
  }%
  \item\relax}
 {\endlist}

\title{Spin in $p$-adic AdS/CFT}

\author{Steven S.~Gubser, Christian Jepsen, and Brian Trundy}
\date{}

\begin{document}
\begin{titlingpage}

\setlength{\droptitle}{-70pt}
\maketitle
\begin{abstract}
We study the holographic dual of the simplest notion of spin in a $p$-adic field theory, namely Green's functions which involve non-trivial sign characters over the $p$-adic numbers.  In order to recover these sign characters from bulk constructions, we find that we must introduce a non-dynamical $U(1)$ gauge field on the line graph of the Bruhat-Tits tree.  Wilson lines of this gauge field on suitable paths yield the desired sign characters.  We show explicitly how to start with complex scalars or fermions in the bulk, coupled to the $U(1)$ gauge field, and compute the holographic two-point functions of their dual operators on the boundary.
\end{abstract}
\vfill
\begin{center}
In memory of Peter G.~O.~Freund
\end{center}
\vfill
February 2019
\end{titlingpage}

\tableofcontents

\section{Introduction}
\label{INTRODUCTION}

Of Peter Freund's many ideas in theoretical physics, it was clear that $p$-adic string theory \cite{Freund:1987kt,Freund:1987ck,brekke1988non}, also studied by Volovich in \cite{Volovich:1987nq}, was one of his favorites.  The strangeness of the $p$-adic numbers, the unexpectedness of Freund and Olson's idea to replace the reals with the $p$-adics on the boundary of the open string worldsheet, and the simplicity of the resulting scattering amplitudes, all contribute to the charm of the subject.  The deep question of why the reciprocal of the Veneziano amplitude factorizes into a product of its $p$-adic relatives remains mysterious.  It causes us to wonder whether, even now, we have fully plumbed the depths of perturbative string dynamics.

In $p$-adic AdS/CFT \cite{Gubser:2016guj,Heydeman:2016ldy}, we are looking at some of the surprising features of $p$-adic string theory in a new light.  In an important precursor to $p$-adic AdS/CFT \cite{Zabrodin:1988ep}, Zabrodin defined a free massless scalar action over the Bruhat-Tits tree $T_p$, whose boundary is the projective line $\mathbb{P}^1(\mathbb{Q}_p)$.  Integrating out the bulk scalar was shown to result in the correlators that Freund and Olson needed to obtain the analog of the Veneziano amplitude for $p$-adic strings.  An updated version of this integrating out process is the computation of holographic Green's functions in $p$-adic AdS/CFT, with the Bruhat-Tits tree playing the role of the bulk geometry.\footnote{See however \cite{Qu:2018ned} for a recent study of holography involving $p$-adic numbers in the context of a continuous bulk geometry.}  Free massless scalars in the bulk are just the beginning: In $p$-adic AdS/CFT one wants to consider mass terms and their relation to boundary conformal dimension, and also non-linear bulk dynamics dual to $n$-point boundary Green's functions with more interesting structure than can be obtained from Wick contractions.

In \cite{Zabrodin:1988ep} as well as later works, attention focused on scalar fields in the bulk geometry, dual to scalar operators on the boundary.  Likewise on the field theory side, the study of the operator product expansion \cite{Melzer1989} focused on scalar operators.  Even the gravitational dynamics of \cite{Gubser:2016htz} is a scalar theory, because the bulk variable is edge length on $T_p$.  The dual boundary operator is found to be a scalar whose scaling dimension equals the dimension of the boundary as a vector space over $\mathbb{Q}_p$.  Boundary theory correlators involving sign characters were considered in \cite{brekke1988non,Arefeva:1988qq,Arefeva:1988qi,Ruelle:1989dg,Marshakov:1989jz} in connection with $p$-adic string amplitudes and supersymmetry.  Work on fermionic $p$-adic field theories continued in \cite{Lerner:1994th} with a study of a relative of the Gross-Neveu model, and the recent work \cite{Gubser:2017qed} investigates both fermionic and bosonic melonic theories over $\mathbb{Q}_p$.  General comments on higher spin can be found in \cite{Heydeman:2016ldy}.  But no bulk dual of non-scalar operators was suggested in any of these works.  Another precursor of $p$-adic AdS/CFT is the stochastic cellular model eternal inflation studied in \cite{Harlow:2011az}; but there too the treatment was restricted to scalar operators on the boundary (best understood as an analog of future infinity in de Sitter space).  In light of the results and suggestions in \cite{Zabrodin:1988ep}, one might say that the question of finding a bulk dual to non-scalar operators has been outstanding for thirty years.  The aim of the current work is to present some first results on this problem.

Taking our cue from \cite{Zabrodin:1988ep,Ruelle:1989dg,Gubser:2017qed}, we consider boundary Green's functions of the form
 \eqn{GreenPropose}{
  G(x) = {C \sgn x \over |x|^{2\Delta}} \,,
 }
where $C$ is a constant, $|\cdot|$ denotes the $p$-adic norm, and $\sgn x$ is a multiplicative sign character on $\mathbb{Q}_p$: that is, a group homomorphism from the multiplicative group $\mathbb{Q}^\times_p$ of non-zero elements of $\mathbb{Q}_p$ to $\{\pm 1\}$.  In \eno{GreenPropose} and below, $|\cdot|$ acting on an element of $\mathbb{Q}_p$ is the $p$-adic norm.  We want to inquire, when and how can we extract a Green's function like \eno{GreenPropose} from a bulk construction?

To further motivate the study of Green's function of the type \eno{GreenPropose}, consider the corresponding Fourier space expression
 \eqn{GreenFourier}{
  \tilde{G}(k) = \tilde{C} (\sgn k) |k|^{2\Delta-1} \,.
 }
For comparison, fermionic correlators in ordinary $\text{AdS}_{d+1}/\text{CFT}_d$ take the form $\tilde{G}(k) = \tilde{C} \gamma_\mu k^\mu |k|^{2\Delta-d-1}$ (where now $|\cdot|$ is the norm on $\mathbb{R}^d$ instead of the $p$-adic norm); see for example \cite{Iqbal:2009fd}.  Our assertion is that the factor $\sgn k$ is in rough analogy to the the factor $\gamma_\mu k^\mu / |k|$ that appears in the Archimedean case.  The first point of similarity is that $(\sgn k)^2 = 1$, just as $(\gamma_\mu k^\mu / |k|)^2 = 1$.  Furthermore, in analogy to the transformations of $\gamma_\mu k^\mu / |k|$ under rotations, $\sgn k$ is a representation of the rotation group on $\mathbb{Q}_p$, which comes from multiplication by $p$-adic numbers with norm $1$.  This is an abelian group, so we only expect to see one-dimensional representations.  There are certainly more complicated representations than just sign characters, so the current work should be considered only a first foray into the potentially large subject of $p$-adic AdS/CFT with spin.

Let's enumerate the sign characters on $\mathbb{Q}_p^\times$ for odd primes $p$.\footnote{Sign characters over $\mathbb{Q}_2$ are also well known, but their relation to holographic constructions is more intricate and will be postponed to to future work.}  First express any $x \in \mathbb{Q}_p^\times$ as
 \eqn{xExpress}{
  x = p^{v_x} (x_0 + x_1 p + x_2 p^2 + \dots) \,,
 }
where $v_x \in \mathbb{Z}$, $x_0 \in \mathbb{F}_p^\times$, and all other $x_i \in \mathbb{F}_p$.  Here $\mathbb{F}_p$ is the finite field of $p$ elements, namely $\mathbb{Z}/p\mathbb{Z}$, which we identify with the set $\{0,1,2,\ldots,p-1\}$.  And $\mathbb{F}_p^\times$ is the non-zero elements of $\mathbb{F}_p$, which form a multiplicative group.  There are two sign characters on $\mathbb{F}_p^\times$: the trivial one which maps all elements to $1$, and the quadratic residue character $n \to (n|p)$ where $(n|p)$, also denoted $\left( n \over p \right)$, is the Legendre symbol.  It is defined so that $(n|p) = 1$ if $n$ is a square in $\mathbb{F}_p^\times$ and $-1$ otherwise.\footnote{The definition of the Legendre symbol is traditionally extended to all of $\mathbb{F}_p$ by defining $(0|p) = 0$, and to all integers by first reducing them modulo $p$.  Then $(n|p)=0$ or $1$ precisely when $n$ is a quadratic residue modulo $p$.}  On $\mathbb{Q}_p^\times$, there are four choices of sign character:
 \begin{enumerate}
  \item We can map all $x \in \mathbb{Q}_p^\times$ to $1$.  This is the trivial character.
  \item We can map $x \to (x_0|p)$.  
  \item We can map $x \to (-1)^{v_x}$.  This means we assign $p$ itself a sign of $-1$.
  \item We can map $x \to (-1)^{v_x} (x_0|p)$.
 \end{enumerate}
This list exhausts all the sign characters on $\mathbb{Q}_p^\times$.  Until we get to section~\ref{OTHERS}, we are going to focus exclusively on the second case: That is, we will hereafter define 
 \eqn{sgnxDef}{
  \sgn x = \left( x_0 \over p \right) \,.
 }
We will narrow our field of inquiry in two other ways.  First, we will restrict attention to nearest neighbor interactions in the bulk, expressible in terms of a classical action either on $T_p$ or on its line graph $L(T_p)$.  This is analogous to restricting to the lowest non-trivial order in derivatives in Archimedean anti-de Sitter space.  Second, the boundary for us will always be $\mathbb{Q}_p$ rather than an extension of $\mathbb{Q}_p$.  We anticipate that our results should be capable of generalization to arbitrary extensions of $\mathbb{Q}_p$.

The organization of the rest of the paper is as follows.  In section~\ref{ACTION}, we describe the nearest neighbor actions on $T_p$ and $L(T_p)$ that we will need, both for bosons and fermions.  In order to obtain a factor of $\sgn x$ in the final holographic two-point functions, we need to introduce a non-dynamical $U(1)$ gauge field.  Indeed, the factor of $\sgn x$ in \eno{GreenPropose} can be thought of as a Wilson line obtained by integrating the $U(1)$ gauge field along the shortest path on $L(T_p)$ between the boundary points $0$ and $x$.  The particular gauge field configurations that we need are described in section~\ref{BACKGROUND}.  The main technical steps in extracting the holographic two-point functions are outlined in section~\ref{PROPAGATOR}, which deals with bulk-to-boundary propagators, and section~\ref{GREEN}, which recounts the holographic prescription.  We detour briefly in section~\ref{GAUGE} into an account of dynamical gauge fields in the bulk, and then in section~\ref{OTHERS} we summarize how to modify the gauge fields so as to get any sign character one wants in the final two-point function \eno{GreenPropose} (for $p \neq 2$).

We were delighted to read in \cite{Freund:2017aqf} Peter Freund's perspective on $p$-adic string theory and $p$-adic AdS/CFT.  He gave us a lot to think about, and we like to think he would have enjoyed this paper.

\section{Nearest neighbor actions}
\label{ACTION}

In this section we will formulate nearest neighbor actions, first for bosons in section~\ref{BOSON} and then for fermions in section~\ref{FERMION}.  A key ingredient will be a non-dynamical $U(1)$ gauge field.

\subsection{Bosonic actions}
\label{BOSON}

Starting with a complex-valued function $\phi_a$ on vertices $a$ of a directed graph, we can define the gradient of $\phi$ as the following complex-valued function on edges of the graph:\footnote{Nothing so far privileges complex numbers: $\phi_a$ and $d\phi_e$ could be valued in any linear space $V$, and then $\omega_e$ as used in \eno{dDaggerDef}-\eno{IntegrationByParts} would need to be valued in linear functions on $V$.  For the most part we do not need such a general viewpoint.}
 \eqn{dDef}{
  d\phi_e = \phi_{t(e)} - \phi_{s(e)} \,.
 }
Here $e$ is an oriented edge with starting point $s(e)$ and terminus $t(e)$.

If we start from a function $\omega_e$ on directed edges, then we define
 \eqn{dDaggerDef}{
  d^T \omega_a = \sum_{t(e) = a} \omega_e - \sum_{s(e) = a} \omega_e \,,
 }
so that
 \eqn{IntegrationByParts}{
  \sum_e \omega_e d\phi_e = \sum_a (d^T \omega_a) \phi_a \,,
 }
possibly up to issues of boundary terms and/or convergence.  The equality \eno{IntegrationByParts} is an analog of integration by parts.  It can be useful to think of $d = d_{ea}$ as a rectangular matrix with one edge-valued index $e$ and one vertex-valued index $a$.  Then, for example, $d\phi_e = \sum_a d_{ea} \phi_a$, and $d^T \omega_a = \sum_e d_{ea} \omega_e$.

\begin{figure}
  \begin{center}
  \begin{tikzpicture}{>=stealth}
   \tikzmath{\pointsize = 0.1;}
   \coordinate (a) at (0,1);
   \coordinate (b) at (0,-1);
   \coordinate (d) at (1.5,0);
   \draw[fill] (a) circle [radius = \pointsize];
   \draw[fill] (b) circle [radius = \pointsize];
   \draw[thick] (a) -- (b);
   \path (a) -- (b) node[midway] (e) {}; \draw[thick,->] (a) -- (e);
   \node[left=2pt,yshift=-4pt] at (a) {$a$};
   \node[left=2pt,yshift=4pt] at (b) {$b$};
   \node[right=2pt,yshift=-4pt] at (a) {$\phi_a$};
   \node[right=2pt,yshift=4pt] at (b) {$\phi_b$};
   \node[left] at (e) {$e$};
   \node[right=2pt,yshift=2.2pt] at (e) {$e^{i\theta_e}$};
   \node[anchor=west] at (d) {$D\phi_e = e^{i\theta_e} \phi_b - \phi_a$};
   \coordinate (a1) at (-1,1.5);
   \draw[fill] (a1) circle [radius = \pointsize];
   \draw[thick] (a) -- (a1);
   \path (a) -- (a1) node[near end] (c) {}; \draw[thick,->] (a) -- (c);
   \coordinate (a2) at (1,1.5);
   \draw[fill] (a2) circle [radius = \pointsize];
   \draw[thick] (a) -- (a2);
   \path (a) -- (a2) node[near end] (c) {}; \draw[thick,->] (a) -- (c);
   \coordinate (b1) at (-1,-1.5);
   \draw[fill] (b1) circle [radius = \pointsize];
   \draw[thick] (b) -- (b1);
   \path (b) -- (b1) node[near end] (c) {}; \draw[thick,->] (b) -- (c);
   \coordinate (b2) at (0,-2);
   \draw[fill] (b2) circle [radius = \pointsize];
   \draw[thick] (b) -- (b2);
   \path (b) -- (b2) node[near start] (c) {}; \draw[thick,->] (b2) -- (c);
   \coordinate (b3) at (1,-1.5);
   \draw[fill] (b3) circle [radius = \pointsize];
   \draw[thick] (b) -- (b3);
   \path (b) -- (b3) node[near start] (c) {}; \draw[thick,->] (b3) -- (c);
  \end{tikzpicture}
  \end{center}
  \caption{The gauge covariant derivative $D\phi_e$ on a small section of a directed graph.}\label{Dphi}
 \end{figure}
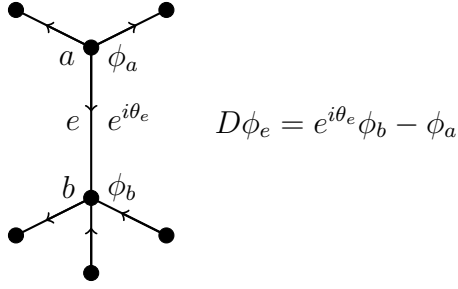

A crucial ingredient in our constructions is a non-dynamical $U(1)$ gauge field.  Because the graph is discrete, instead of a gauge-covariant derivative $D_\mu = \partial_\mu + i A_\mu$, we are going to consider modifying \eno{dDef} to
 \eqn{DphiDef}{
  D\phi_e = e^{i\theta_e} \phi_{t(e)} - \phi_{s(e)} \,,
 }
where $\theta_e$ is essentially $\int A_\mu dx^\mu$ across the edge $e$.  See figure~\ref{Dphi}.
Upon gauge transformations
 \eqn{Uone}{
  \phi_a \to e^{i\lambda_a} \phi_a \qquad\qquad
   \theta_e \to \theta_e - d\lambda_e \,,
 }
we see that
 \eqn{DphiTrans}{
  D\phi_e \to e^{i\lambda_{s(e)}} D\phi_e \,.
 }
Evidently, in the absence of loops, we can use \eno{Uone} with $\lambda_e = -\theta_e$ to remove the phase from \eno{DphiDef}, so that $D = d$.  On $L(T_p)$, there are loops, so non-trivial gauge field configurations exist.

We now consider the action
 \eqn{ScalarAction}{
  S_\phi = \sum_e |D\phi_e|^2 + \sum_a m^2 |\phi_a|^2 \,.
 }
Here and below, $|\cdot|$ acting on a target space field is the norm on $\mathbb{C}$, whereas $|\cdot|$ acting on coordinates or momenta is the $p$-adic norm.  Varying \eno{ScalarAction} with respect to $\phi_a^*$ gives
 \eqn{EomPhi}{
  D^\dagger D \phi_a + m^2 \phi_a = 0 \,,
 }
where $D^\dagger$ is the adjoint of $D$.  (Explicitly, we can write $D = D_{ea}$ as a rectangular matrix, and then $D^*$, $D^T$, and $D^\dagger$ all have obvious definitions.)  A helpful result for calculations to come is
 \eqn{DDlemma}{
  D^\dagger D \phi_a = o_a \phi_a - \sum_{t(e)=a} e^{-i\theta_e} \phi_{s(e)} - 
    \sum_{s(e)=a} e^{i\theta_e} \phi_{t(e)} \,,
 }
where $o_a$ is the number of edges incident upon $a$.  (On $T_p$, $o_a = p+1$ for all vertices, while on $L(T_p)$, $o_a = 2p$ for all vertices.)

The definition \eno{DphiDef} might seem asymmetrical, and one might prefer instead $e^{i\theta_e/2} \phi_{t(e)} - e^{-i\theta_e/2} \phi_{s(e)}$ on the right hand side.  But for purposes of forming the action \eno{ScalarAction}, the overall phase of $D\phi_e$ doesn't matter because only $|D\phi_e|^2$ enters.  In other words, there is $U(1)$ gauge freedom on edges which we fix in \eno{DphiDef} by locking the phase of $D\phi_e$ to $\phi_{s(e)}$.

\subsection{Fermionic actions}
\label{FERMION}

On a graph, the natural notion of a Dirac operator has to do with the exterior derivative.  See for example \cite{Knill:2013zz}, where the Dirac operator on a graph $D$ is essentially the signed adjacency matrix on the clique graph of $G$.  We will consider a simplification of this general development, in which the only operator we need is the gradient, rendered gauge covariant as in the previous section.  Explicitly, we introduce a Grassmann-complex-valued function $\psi_a$ on vertices of a directed graph, and another such function $\chi_e$ on edges.  We define
 \eqn{DpsiDef}{
  D\psi_e = e^{i\theta_e} \psi_{t(e)} - \psi_{s(e)}
 }
and introduce the action
 \eqn{Sfermion}{
  S_\psi = 
   \sum_e \left[ i \chi_e^* D\psi_e + i \chi_e D^*\psi_e^* + m \chi_e^* \chi_e \right] - 
    \sum_a M \psi_a^* \psi_a \,.
 }
The kinetic terms in \eno{Sfermion} are constructed in the spirit of $b \partial c$ lagrangians, where $b$ is replaced by an edge field $\chi_e$ and $c$ is replaced by a vertex field $\psi_a$.  The action is real once we assume that conjugation exchanges the order of factors.  We need $\psi_a$ and $\chi_e$ to be complex in order to make the mass terms possible.  It appears that the two mass coefficients are independently meaningful, but in fact there is a global scaling symmetry $\psi_a \to \lambda\psi_a$ and $\chi_e \to (\lambda^*)^{-1}\chi_e$, where $\lambda \in \mathbb{C}$ is a constant, which preserves the kinetic terms while rescaling $m \to |\lambda|^2 m$ and $M \to |\lambda|^{-2} M$.

The action \eno{Sfermion} is invariant under the gauge transformation
 \eqn{UonePsi}{
  \psi_a \to e^{i\lambda_a} \psi_a \qquad\qquad 
  \chi_e \to e^{i\lambda_{s(e)}} \chi_e \qquad\qquad
  \theta_e \to \theta_e - d\lambda_e \,,
 }
and the equations of motion are
 \eqn{EomsFermion}{
  iD \psi_e + m \chi_e = 0 \qquad\qquad
   i D^\dagger \chi_a + M \psi_a = 0
 }
(and the complex conjugates of these equations).\footnote{As in the scalar case, the overall phase of $D\psi_e$ doesn't matter because in \eno{Sfermion} we form the product $i\chi_e^* D\psi_e$, and we can adjust the phase of $\chi_e$ to keep the overall prefactor equal to $i$.}  From the two equations \eno{EomsFermion} it follows that
 \eqn{EomPsi}{
  d^\dagger d \psi_a + mM \psi_a = 0 \,.
 }
The equivalence of \eno{EomPsi} and \eno{EomPhi} is comparable to the way the massive Dirac equation implies the massive Klein-Gordon equation.

\section{The background geometries}
\label{BACKGROUND}

The non-dynamical $U(1)$ gauge fields on $L(T_p)$ that we are going to study encode the Legendre symbol $(\alpha|p)$.  Consider first the case $p \equiv 1 \mod 4$.  Label the vertices of the complete graph $K_p$ with elements of $\mathbb{F}_p$.  Pick a directed structure on $K_p$, and define a map 
 \eqn{alphaMapDef}{
  e \to \alpha(e) = t(e)-s(e)
 }
from directed edges to $\mathbb{F}_p^\times$.  Set $e^{i\theta_e} = (\alpha(e)|p)$ on each edge.  Because $(\alpha|p)$ is an even function of $\alpha \in \mathbb{F}_p^\times$, the choice of $e^{i\theta_e}$ doesn't depend on the directed structure we picked.  Because $e^{i\theta_e}$ is always real, the operator $D^\dagger D$ also doesn't depend on the directed structure.

Now consider the case $p \equiv 3 \mod 4$.  Again label the vertices of $K_p$ with elements of $\mathbb{F}_p$.  Introduce a directed structure on $K_p$ such that an edge runs from a vertex $a$ to another vertex $b$ iff $b-a$ is a square in $\mathbb{F}_p^\times$.  This prescription uniquely specifies the direction of every edge in $K_p$ because for any $\alpha \in \mathbb{F}_p^\times$, either $(\alpha|p) = 1$ or else $(-\alpha|p) = 1$, due to the fact that $(\alpha|p)$ is an odd function of $p$.  Set $e^{i\theta_e} = i$ on all edges.

We will refer to the directed structures and gauge fields on $K_p$ as Paley constructions, since for $p \equiv 1 \mod 4$ the edges with $e^{i\theta_e} = 1$ form a Paley graph (without reference to the directed structure), while for $p \equiv 3 \mod 4$, the directed structure that we picked forms a Paley digraph.

There are $p+1$ edges incident upon each vertex $A$ of $T_p$, of which one edge is below located below $A$ (that is, one edge lies on the path from $A$ to the boundary point at infinity), while $p$ edges are located above $A$. The vertices in $L(T_p)$ corresponding to the above-lying edges we think of as forming a copy of $K_p$, and each of these vertices is also connected to the vertex corresponding to the edge below $A$.  Following \cite{Gubser:2016guj}, we parametrize $A$ using a pair $(x_A,z_A)$ where $x_A \in \mathbb{Q}_p$ and $z_A$ is an integer power of $p$, with $(x_A,z_A)$ identified with $(x'_A,z'_A)$ iff $z_A=z'_A$ and $|x_A-x'_A| \leq |z_A|$, where $|z_A|$ indicates the $p$-adic norm of $z_A$.  We will parametrize elements of $L(T_p)$ by using the same coordinates $(x_A,z_A)$ to label the vertex of $L(T_p)$ immediately below $A$; usually we will write instead $(x_a,z_a)$ since we use lowercase letters to label vertices of $L(T_p)$.  To fix a directed structure and non-dynamical gauge field on $L(T_p)$, we adopt the same Paley construction on each $K_p$, and the rest of the edges are directed downward, from $(x,z)$ to $(x,z/p)$, with $e^{i\theta_e} = 1$ for $p \equiv 1 \mod 4$ and $e^{i\theta_e} = i$ for $p \equiv 3 \mod 4$.  We will refer to edges inside a copy of $K_p$ as horizontal, and the others as vertical.

\begin{figure}
\centering{
\includegraphics[height=34ex]{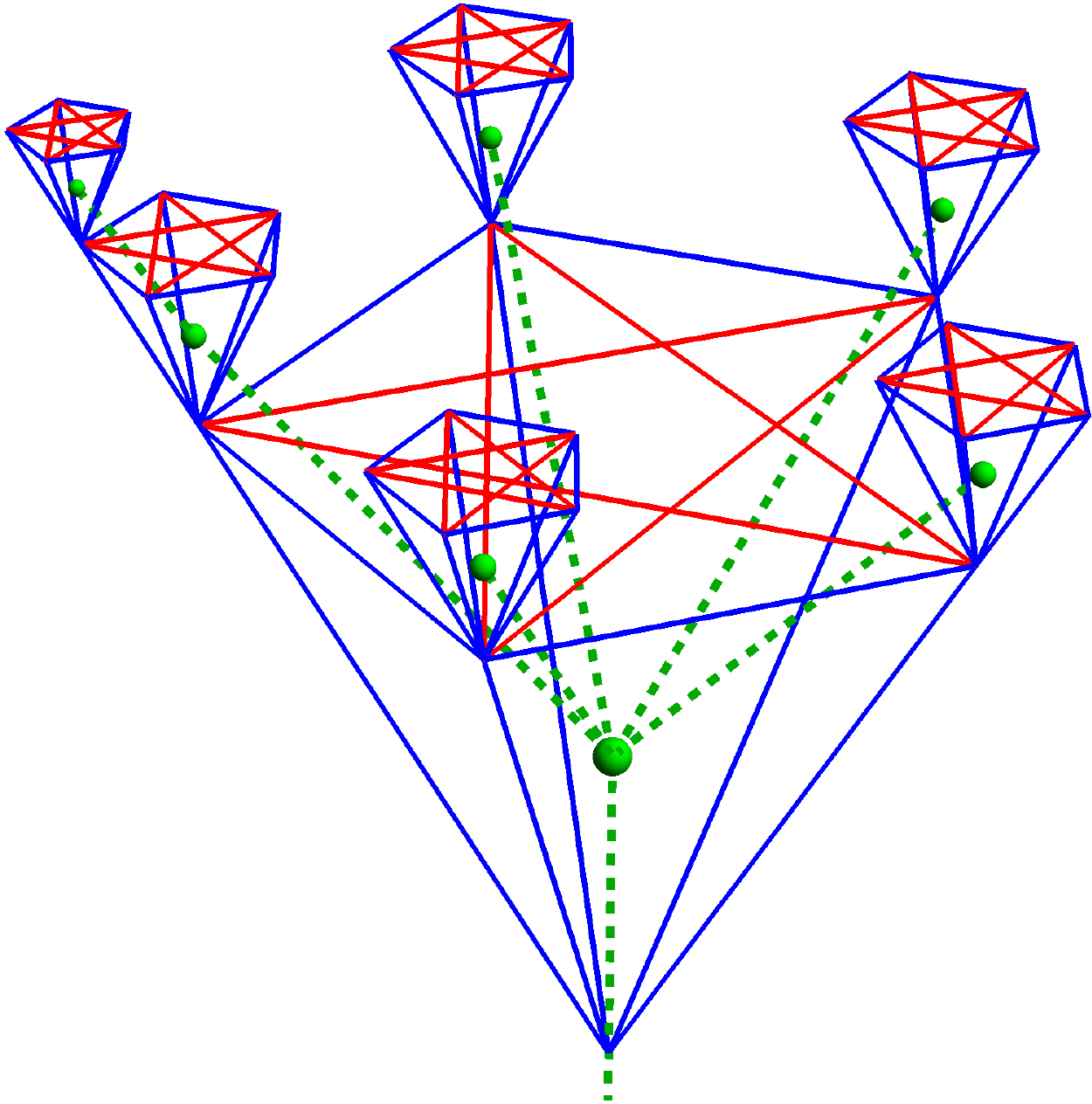} \hspace{5em}
\includegraphics[height=34ex]{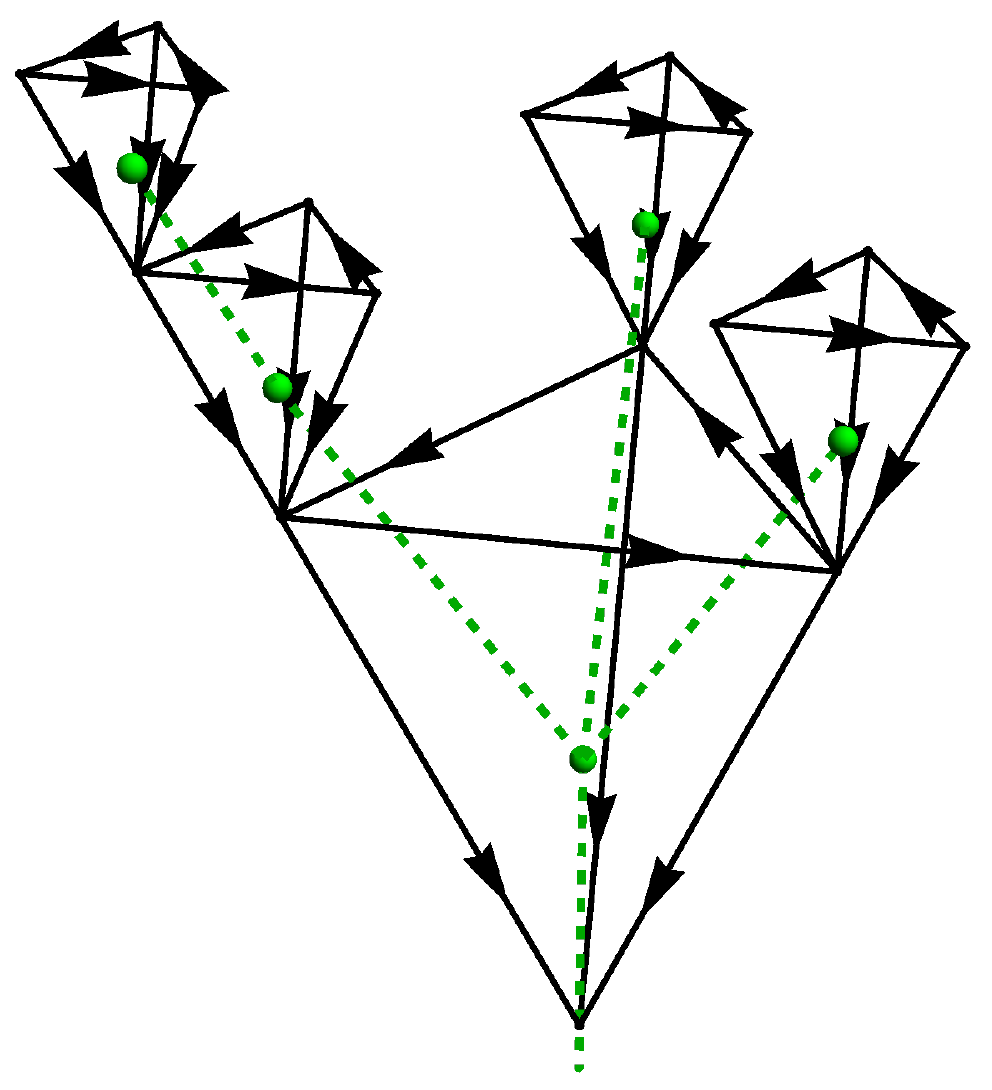}
}
\caption{Left: $L(T_5)$ in blue and red with $T_5$ shown in dashed green.  The blue edges have $\theta_e = 0$, while the red edges have $\theta_e = \pi$.  The pentagram figures are the Paley constructions, and edges within them are called horizontal.  Their orientation doesn't matter.  The vertical edges are the ones connecting each pentagram with a vertex of the pentagram below it, and our convention is for all of them to be directed downward.\\  Right: $L(T_3)$ in black, with $T_3$ shown in dashed green.  All the edges have $\theta_e = \pi/2$.}
\label{LineGraphs}
\end{figure}

If we use $a \to b$ to denote a directed edge, then for $p\equiv 1 \mod 4$, our choice of gauge field is
 \eqn{FiveSummary}{\seqalign{\span\TT &\qquad \span\TL & \span\TR &\qquad\span\TT}{
  vertical edges: & \theta_{(x,z) \to \left( x,{z \over p} \right)} &= 0  \cr
  horizontal edges: & 
   \theta_{(x,z) \to \left( x + \alpha {z \over p}, z\right)} &= 
     {\pi \over 2} \left[ 1 - \left( \alpha \over p \right) \right] &
     for $\alpha \in \mathbb{F}_p^\times$\,.
 }}
In the second line of \eno{FiveSummary}, we bear in mind that the directed edge $(x,z) \to \left( x + \alpha {z \over p}, z\right)$ exists only for half the elements of $\mathbb{F}_p^\times$; but which half doesn't matter.  For $p\equiv 3 \mod 4$, our choice of gauge field is
 \eqn{ThreeSummary}{\seqalign{\span\TT &\qquad \span\TL & \span\TR &\qquad\span\TT}{
  vertical edges: & \theta_{(x,z) \to \left( x,{z \over p} \right)} &= {\pi \over 2}  \cr
  horizontal edges: & 
   \theta_{(x,z) \to \left( x + \alpha {z \over p}, z\right)} &= 
     {\pi \over 2} &
     for $\alpha \in \mathbb{F}_p^\times$ with $\left( \alpha \over p \right) = 1$\,.
 }}
See figure~\ref{LineGraphs} for a depiction of small subgraphs of $L(T_p)$ showing also the choice of gauge fields \eno{FiveSummary} and \eno{ThreeSummary} for $p=5$ and $3$, respectively.

The assignments of $e^{i\theta_e}$ in \eno{FiveSummary} and \eno{ThreeSummary} are preserved under the maps $(x,z) \to (rx+b,z/|r|)$ for all $b \in \mathbb{Q}_p$ and $r \in (\mathbb{Q}_p^\times)^2$.  This is fortunate because the corresponding boundary maps, $x \to rx + b$, applied to two distinct points $x_1$ and $x_2$ in $\mathbb{Q}_p$, are the ones that preserve the desired two-point function, $G(x_1,x_2) = \sgn (x_1-x_2)/|x_1-x_2|^{2\Delta}$, up to some power of the scale factor $|r|$.

\section{Bulk-to-boundary propagators}
\label{PROPAGATOR}

In order to compute holographic two-point functions, a key ingredient is the bulk-to-boundary propagator in momentum space.  In section~\ref{TREE} we review the calculation of this propagator in the case of complex scalars on $T_p$.  Then in section~\ref{LINE} we work it out for scalars and fermions on $L(T_p)$.

\subsection{Scalars on \texorpdfstring{$T_p$}{Tp}}
\label{TREE}

Consider complex scalars on $T_p$ with action
 \eqn{ScalarTp}{
  S = \sum_E |d\phi_E|^2 + \sum_A m^2 |\phi_A|^2 \,,
 }
where vertices are labeled $A = (x_A,z_A)$ and edges are labeled $E$.  For brevity, let's write $z_A = p^v$.  Then, as recounted in \cite{Gubser:2016guj}, a useful solution of the equations of motion following from \eno{ScalarTp} is
 \eqn{KakSoln}{
  \phi_A(k) = f_v \gamma(kz_A) \chi(-kx_A) \qquad\text{where}\qquad 
   f_v = |z_A|^{1-\Delta} + Q |k|^{2\Delta-1} |z_A|^\Delta
 }
where $k \in \mathbb{Q}^\times_p$ and
 \eqn{Qvalue}{
  Q = -p^{1-2\Delta} \,.
 }
Here $\chi(\xi)$ is the additive character on $\mathbb{Q}_p$, given explicitly by $\chi(\xi) = e^{2\pi i \{\xi\}}$ where $\{\xi\}$ is the fractional part of $\xi \in \mathbb{Q}_p$.  We define
 \eqn{GammaDef}{
  \gamma(\xi) = \left\{ \seqalign{\span\TL &\qquad\span\TT}{
   1 & for $|\xi| \leq 1$  \cr  0 & otherwise\,,} \right.
 }
and
 \eqn{ZetaDef}{
  \zeta(s) = {1 \over 1 - p^{-s}} \,.
 }
The dimension $\Delta$ is related to the mass by
 \eqn{MassDim}{
  m^2 = -{1 \over \zeta(\Delta-1) \zeta(-\Delta)} \,.
 }
The solution $\phi_A(k)$ in \eno{KakSoln} can be thought of as a bulk to boundary propagator because it is the disturbance of $\phi_A$ in the bulk that corresponds to deforming the boundary field theory by a term $\int dx \, \chi(kx) {\cal O}(x)$, where ${\cal O}$ is the operator dual to $\phi$.\footnote{The field $\phi_a$, the operator ${\cal O}$ and the deformation of the conformal field theory action, should in the end be real.  This can be accomplished by always considering superpositions of Fourier modes with equal and opposite $k$.}  From the form of $\phi_A(k)$ given in \eno{KakSoln} we can pick out the $k$-dependence of the Fourier space holographic two-point function for ${\cal O}$: $\tilde{G}(k) \propto |k|^{2\Delta-1}$.

\subsection{Scalars and fermions on the line graph}
\label{LINE}

We would now like to find solutions analogous to \eno{KakSoln} on $L(T_p)$ with the directed structures and gauge fields as outlined in section~\ref{BACKGROUND}.  The invariance of the background geometry under translations $x_a \to x_a + b$ indicates that we should be able to require that fields should depend on $x_a$ through a factor $\chi(-kx_a)$.  We immediately encounter the need to multiply in a factor of $\gamma(kz_a)$, because by itself, $\chi(-kx_a)$ is not single valued on $L(T_p)$, whereas $\gamma(kz_a) \chi(-kx_a)$ is (and the same logic dictated that \eno{KakSoln} must include a factor of $\gamma(k z_A)$).  In short, we are lead to essentially the same ansatz as \eno{KakSoln}:
 \eqn{KakAnsatz}{
  \phi_a(k) = u^v f_v \gamma(kz_a) \chi(-kx_a) \qquad\text{where}\qquad
    f_v = |z_a|^{1-\Delta} + Q |k|^{2\Delta-1} |z_a|^\Delta
 }
and we have written $z_a = p^v$.  For later convenience, we have introduced the prefactor $u^v$ where
 \eqn{uValues}{
  u = \left\{ \seqalign{\span\TR &\qquad \span\TT}{
    1 & for $p \equiv 1 \mod 4$  \cr
    i & for $p \equiv 3 \mod 4$\,.
   } \right.
 }
The coefficient $Q$ in \eno{KakAnsatz} may depend on $k$, but we don't expect it to depend on $|k|$ since the explicit factor of $|k|^{2\Delta-1}$ already is the dependence we expect for the $k$-th Fourier mode of a holographic two-point function with dimension $\Delta$.  The aims of the following calculation are to verify that the ansatz \eno{KakAnsatz} does solve \eno{EomPhi} and to determine $\Delta$ and $Q$.  The strategy is to plug \eno{KakAnsatz} into \eno{EomPhi} and extract a difference equation for $f_v$.  We can handle the fermionic case by replacing $\phi_a$ by $\psi_a$ in \eno{KakAnsatz} and plugging into \eno{EomPsi}.  In the fermionic case, we assume $m \neq 0$ so that $\chi_e$ can be determined by the first equation in \eno{EomsFermion}.

The factor $\gamma(kz_a)$ means that the equation of motion \eno{EomPhi} is trivially satisfied for vertices $a$ such that $v_k + v < -1$, where $v_k \in \mathbb{Z}$ is the valuation of $k$: that is
 \eqn{kForm}{
  k = p^{v_k} (k_0 + k_1 p + k_2 p^2 + \dots) 
   \qquad\text{with $k_0 \in \mathbb{F}_p^\times$}\,.
 }
Let's first show that the equation of motion is also trivially satisfied when $v = -v_k-1$.  In this case, the only non-zero terms in the equation of motion \eno{EomPhi} are the ones corresponding to $z_b = p z_a$, i.e.~in the Paley construction above the point $a$.  Explicitly, the equation of motion reads
 \eqn{fvMinusOne}{
  -f_{-v_k} \sum_{\alpha \in \mathbb{F}_p} \chi(-k[x_a + z_a \alpha]) = 0 \,,
 }
where we are using the fact that the $p$ vertices in $L(T_p)$ above $(x_a,z_a)$ are $(x_a+z_a \alpha,pz_a)$ where $\alpha$ runs over $\mathbb{F}_p$.  Recalling that $v = -v_k-1$, we see that the sum in \eno{fvMinusOne} is proportional to
 \eqn{chiTrivial}{
  \sum_{\alpha \in \mathbb{F}_p} \chi(-k_0 \alpha/p) = 0 \,.
 }
So \eno{fvMinusOne} is indeed satisfied trivially and gives us no information about $f_v$.

Let's move on to the case $v > -v_k$.  The factor $\chi(-kx_b)$ now has the same value for all vertices $b$ neighboring $a$, as well as for $b=a$.  Also, $\gamma(kz_b) = 1$ at all these vertices.  Thus we may discard the factor $\gamma(kz_a) \chi(-kx_a)$ from the ansatz \eno{KakAnsatz} and work directly with $\phi_a = f_v$.  By plugging in to \eno{EomPhi} we find
 \eqn{fvRecursion}{
  (2p+m^2) f_v - p f_{v+1} - f_{v-1} = 0 \,.
 }
The form of \eno{fvRecursion} is the same for $p\equiv 1 \mod 4$ and $p\equiv 3 \mod 4$ because of the overall prefactor $u^v$ in \eno{KakAnsatz}.  One can view this factor as a change of gauge in the $p\equiv 3 \mod 4$ case which removes the factors of $i$ from the covariant derivatives along vertical edges while leaving them unchanged within the Paley constructions.  The only property of the Paley constructions we need in order to get \eno{fvRecursion} is that contributions to $D^\dagger D \phi_a$ from the $p-1$ vertices connected to $a$ by a horizontal edge cancel out.  The second order difference equation \eno{fvRecursion} is solved by $f_v = p^{-\Delta v} = |z_a|^\Delta$ and $f_v = p^{(\Delta-1) v} = |z_a|^{1-\Delta}$ where
 \eqn{DeltaMrelation}{
  2p+m^2 = p^{1-\Delta} + p^\Delta \,.
 }
We assume that $\Delta$ is real, and the standard prescription is to choose it as the larger of the two roots of \eno{DeltaMrelation}, so that $\Delta > 1/2$.  Note that $2(\sqrt{p}-p) < m^2 < 0$ when $\Delta \in (1/2,\log_p(p+\sqrt{p(p-1)}))$, and it is positive otherwise.\footnote{Note that $\Delta = {1 \over 2} + is$ gives $m^2$ real but violating the lower bound $m^2 > 2(\sqrt{p}-p)$.  A similar result was already noted in \cite{Heydeman:2016ldy} for scalars on $T_p$.  It is tempting to think that these complex values of $\Delta$ correspond to unstable actions, but they may nevertheless have some interesting role to play.}

In the case of fermions, the discussion up to this point proceeds unchanged, except that $m^2$ is replaced by $mM$.

To summarize progress so far: We have shown that the ansatz \eno{KakAnsatz} trivially satisfies the equations of motion for $v < -v_k$, while for $v > -v_k$ we have shown that it is consistent with the equations of motion provided we impose the mass-dimension relation \eno{DeltaMrelation}.  But we have no information yet about $Q$.  This information comes from a boundary condition at $v = -v_k$, and it turns out that it encodes the sign character that we need in order to obtain two-point functions of the desired form \eno{sgnxDef}.  The equation of motion \eno{EomPhi} for $v=-v_k$ reads
 \eqn{fvBoundary}{
  (2p + m^2) f_{-v_k} \chi(-kx_a) - 
    f_{-v_k+1} \sum_{\alpha \in \mathbb{F}_p} &\chi(-k[x_a + z_a \alpha])  \cr
     &{} - u f_{-v_k} \sum_{\alpha \in \mathbb{F}_p^\times} \left( {\alpha \over p} \right) 
     \chi\left( -k \left[ x_a + z_a {\alpha \over p} \right] \right) = 0 \,.
 }
There is no $f_{-v_k-1}$ term in \eno{fvBoundary} because the factor of $\gamma(kz_b)$ vanishes when $z_b = p^{-v_k-1}$, so \eno{fvBoundary} is only first order in differences rather than second order.  Hence it can indeed be thought of as a boundary condition for the second order equation \eno{fvRecursion}.  In the last term of \eno{fvBoundary}, we are using the fact that the $p-1$ vertices in the same Paley construction as $x_a$ are $\left( x_a + z_a {\alpha \over p},z_a \right)$.  This last term is proportional to the Gauss sum:
 \eqn{GaussSum}{
  \sum_{\alpha \in \mathbb{F}_p} \left( {\alpha \over p} \right) 
    \chi\left( -{k_0 \alpha \over p} \right) = {\sqrt{p} \over u}
     \left( {k_0 \over p} \right) \,.
 }
(In \eno{GaussSum}, the $\alpha=0$ term in the sum vanishes, so including it is optional.  The form \eno{GaussSum} makes it clear that we are taking a Fourier transform of the Legendre symbol over $\mathbb{F}_p$.)  Simplifying, and using \eno{DeltaMrelation}, we obtain
 \eqn{fvSimpler}{
  \left[ p^{1-\Delta} + p^\Delta - \sqrt{p} \left( {k_0 \over p} \right) 
    \right] f_{-v_k} - p f_{-v_k+1} = 0 \,.
 }
Plugging the ansatz for $f_v$ in \eno{KakAnsatz} into \eno{fvSimpler}, one arrives at
 \eqn{Qnontrivial}{
  \left[ p^{1-\Delta} + p^\Delta - \sqrt{p} \left( {k_0 \over p} \right) \right] (1+Q) - 
   p (p^{\Delta-1} + p^{-\Delta} Q) = 0 \,,
 }
which reduces to
 \eqn{Qtrivial}{
  Q = p^{{1 \over 2} - \Delta} \left( {k_0 \over p} \right) 
    = p^{{1 \over 2} - \Delta} \sgn k \,.
 }

\section{Two-point functions}
\label{GREEN}

With the bulk-to-boundary propagators in hand, we now turn to the computation of the holographic two-point functions, first in section~\ref{GREENTREE} for real scalars on $T_p$ and then in section~\ref{GREENLINE} for complex scalars on $L(T_p)$, and finally in section~\ref{GREENPSI} for fermions on $L(T_p)$.

\subsection{Scalars on \texorpdfstring{$T_p$}{Tp}}
\label{GREENTREE}

As a warmup, consider a complex scalar $\phi_A$ on $T_p$, as in \eno{ScalarTp}-\eno{MassDim}.  Implement a cutoff by fixing the values of $\phi_A$ for all vertices $(x_A,z_A)$ with $|z_A| = |\epsilon|$, where $\epsilon = p^{v_\epsilon}$ and $v_\epsilon$ is an integer.  Let $\Sigma_\epsilon$ denote the set of vertices with $|z_A| > |\epsilon|$, together with the edges with at least one vertex having $|z_A| > |\epsilon|$.  Let $\partial \Sigma_\epsilon$ be the edges with only one vertex in $\Sigma_\epsilon$.  We orient edges downward (away from the $\mathbb{Q}_p$ boundary), so that when $E \in \partial \Sigma_\epsilon$, $t(E) \in \Sigma_\epsilon$ and $s(E) \not\in \Sigma_\epsilon$.  The vertices in $\Sigma_\epsilon$ are allowed to fluctuate, while vertices with $|z_A| < |\epsilon|$ are ignored.  The cutoff action is 
 \eqn{CutoffAction}{
  S_\epsilon = \sum_{E \in \Sigma_\epsilon} |d\phi_E|^2 + 
    \sum_{A \in \Sigma_\epsilon} m^2 |\phi_A|^2 \,.
 }
We now need an improvement of the partial integration formula \eno{IntegrationByParts} to include boundary terms:\footnote{If $\partial\Sigma_\epsilon$ included edges for which $s(E) \in \Sigma_\epsilon$ while $t(E) \not\in \Sigma_\epsilon$, then in place of \eno{PartsBoundary} we would need $\sum_{E \in \Sigma_\epsilon} \omega_E d\phi_E = 
   \sum_{A \in \Sigma_\epsilon} (d^T \omega_A) \phi_A +
   \sum_{E \in \partial\Sigma_\epsilon \atop t(E) \not\in \Sigma_\epsilon} 
     \omega_E \phi_{t(E)} -
   \sum_{E \in \partial\Sigma_\epsilon \atop s(E) \not\in \Sigma_\epsilon} 
     \omega_E \phi_{s(E)}$.}
 \eqn{PartsBoundary}{
  \sum_{E \in \Sigma_\epsilon} \omega_E d\phi_E = 
   \sum_{A \in \Sigma_\epsilon} (d^T \omega_A) \phi_A -
   \sum_{E \in \partial\Sigma_\epsilon} 
     \omega_E \phi_{s(E)} \,.
 }
Using \eno{PartsBoundary} we see that
 \eqn{SeBoundary}{
  S_\epsilon &= {1 \over 2} \sum_{A \in \Sigma_\epsilon} \left[ 
    \phi_A^* (d^\dagger d \phi_A + m^2 \phi_A) + 
    (d^T d^* \phi_A^* + m^2 \phi_A^*) \phi_A \right]  \cr
   &\qquad{} - 
    {1 \over 2} \sum_{E \in \partial\Sigma_\epsilon} \left[ 
     \phi_{s(E)}^* d\phi_E + (d\phi_E^*) \phi_{s(E)} \right] \,.
 }
The first line of \eno{SeBoundary} vanishes on-shell, leaving only the boundary terms.  Recalling that $-S_\epsilon^\text{on-shell}$ is the generating function of connected Green's functions and following the logic of \cite{Iqbal:2009fd}, we see that a cutoff version of the Green's function can be computed as
 \eqn{FoundGtilde}{
  \tilde{G}_\epsilon(k) = {d\phi_E(k) \over \phi_{s(E)}(k)} \qquad
   \text{for any $E \in \partial\Sigma_\epsilon$} \,,
 }
where, crucially, we have plugged in the solution $\phi_A = \phi_A(k)$ from \eno{KakSoln}.  We have to choose $|k\epsilon| < 1$ in order to avoid having a vanishing denominator in \eno{FoundGtilde}.  As long as we work at fixed $k$, this is not a problem, since our eventual aim is to take $\epsilon \to 0$ $p$-adically.  Straightforward calculation of the right hand side of \eno{FoundGtilde} gives
 \eqn{CalculateGTilde}{
  \tilde{G}_\epsilon(k) = {f_{v_\epsilon-1} - f_{v_\epsilon} \over f_{v_\epsilon}}
    = -{1 \over \zeta(\Delta-1)} + 
    |k|^{2\Delta-1} |\epsilon|^{2\Delta-1} {Q p^\Delta \over \zeta(2\Delta-1)} +
     \dots \,,
 }
where to obtain the first equality we used \eno{KakSoln}, and to obtain the second we expanded in $p$-adically small $\epsilon$.  The omitted terms go to $0$ more quickly than the ones shown provided $\Delta>1/2$, which is true of the larger of the two roots of the relation \eno{MassDim}.  The first term in \eno{CalculateGTilde} is $k$-independent, so in position space it gives rise to a pure contact term.  Dropping this term, we define the Fourier space Green's function as
 \eqn{GRen}{
  \tilde{G}(k) = 
    \lim_{\epsilon \to 0} {\tilde{G}_\epsilon(k) \over |\epsilon|^{2\Delta-1}}
   = {Q p^\Delta \over \zeta(2\Delta-1)} |k|^{2\Delta-1} \,.
 }
Plugging in $Q = -p^{1-2\Delta}$ from \eno{Qvalue} and recalling the Fourier transform\footnote{An exposition of of Fourier integrals including \eno{BasicFourier} can be found in \cite{Gubser:2017qed}.}
 \eqn{BasicFourier}{
  \int_{\mathbb{Q}_p} dk \, \chi(kx) |k|^s = {\zeta(1+s)/\zeta(-s) \over |x|^{1+s}} \,,
 }
we obtain
 \eqn{GxTp}{
  G(x) = {p^\Delta \zeta(2\Delta) \over \zeta(2\Delta-1)^2} {1 \over |x|^{2\Delta}} \,.
 }
A somewhat more involved derivation of \eno{GxTp} in \cite{Gubser:2016guj} makes it clear that the overall normalization of $G(x)$ is a subtle issue.  Changing the location of the cutoff by one lattice spacing results in changing $G(x)$ by an $O(1)$ multiplicative factor.  We should in short view \eno{GRen} as a reasonable but non-unique prescription for normalizing the two-point function.

\subsection{Scalars on the line graph}
\label{GREENLINE}

For complex scalars $\phi_a$ on $L(T_p)$, the extraction of a Green's function from the bulk to boundary propagator \eno{KakAnsatz} proceeds almost exactly as in the warmup exercise outlined in the previous section.  Formally, in \eno{CutoffAction}-\eno{GRen}, one replaces $d \to D$, $A \to a$, and $E \to e$.  Let's inquire a little more closely why this works.  The set $\Sigma_\epsilon$ comprises vertices with $|z_a| > |\epsilon|$ and edges with at least one vertex having $|z_a| > |\epsilon|$.  The boundary $\partial \Sigma_\epsilon$ consists of vertical edges only, and these edges all have $s(e) \not\in \Sigma_\epsilon$.  Thus the partial integration formula \eno{PartsBoundary} can indeed be carried over to scalars on $L(T_p)$ just by replacing $d \to D$, $A \to a$, and $E \to e$.  Likewise, the subsequent manipulation of the action in \eno{SeBoundary} and the formula \eno{FoundGtilde} for the Green's function carry over with the same alterations.  The calculation \eno{CalculateGTilde} carries over unaltered because of our careful inclusion of a factor of $u^v$ in the scalar ansatz \eno{KakAnsatz}; a more conceptual way to say it is that this factor brings us to a gauge where $D=d$ on vertical edges.  The result \eno{GRen} carries over unaltered, and if we plug in $Q = p^{{1 \over 2} - \Delta} \sgn k$, as given in \eno{Qtrivial}, we obtain
 \eqn{GkSpin}{
  \tilde{G}(k) = {\sqrt{p} \sgn k \over \zeta(2\Delta-1)} |k|^{2\Delta-1} \,.
 }
Using the Fourier integral
 \eqn{pFiveFourier}{
  \int_{\mathbb{Q}_p} dk \, 
    \chi(kx) |k|^s \sgn k = u p^{s+{1 \over 2}} {\sgn x \over |x|^{s+1}} \,,
 }
we arrive at
 \eqn{Gx}{
  G(x) = {u p^{2\Delta} \over \zeta(2\Delta-1)} {\sgn x \over |x|^{2\Delta}} \,.
 }
In terms of the boundary field theory, $G(x) = \langle {\cal O}_\phi(x) {\cal O}_\phi(0)^\dagger \rangle$ where ${\cal O}_\phi$ is the operator dual to $\phi$.  If we assume that translation by $x$ is implemented in the boundary theory by a unitary operator $U(x)$, then ${\cal O}_\psi(x) = U(x)^\dagger {\cal O}_\psi(0) U(x)$, and
 \eqn{GxConj}{
  G(x)^* = \langle {\cal O}_\phi(0) {\cal O}_\phi(x)^\dagger \rangle =
   \langle {\cal O}_\phi(-x) {\cal O}_\phi(0)^\dagger \rangle = G(-x) \,.
 }
The relation $G(x)^* = G(-x)$ is indeed satisfied by \eno{Gx}: For $p\equiv 1 \mod 4$, $G(x)$ is real and even under $x \to -x$, while for $p\equiv 3 \mod 4$, it is imaginary and odd under $x \to -x$.

\subsection{Fermions on the line graph}
\label{GREENPSI}

To derive a holographic two-point function for the operator ${\cal O}_\psi$ dual to $\psi$, we start from the cutoff action
 \eqn{Scutoff}{
  S_\epsilon = \sum_{e \in \Sigma_\epsilon} [ i \chi_e^* D\psi_e + 
    i \chi_e D^* \psi_e^* + m \chi_e^* \chi_e] - 
    \sum_{a \in \Sigma_\epsilon} M \psi_a^* \psi_a \,,
 }
where $\Sigma_\epsilon$ is defined as in section~\ref{BOSON}.  After integration by parts,
 \eqn{Ssym}{
  S_\epsilon &= \sum_{e \in \Sigma_\epsilon} \left[
    {1 \over 2} \chi_e^* (i D \psi_e + m\chi_e) + 
    {1 \over 2} \chi_e (i D^* \psi_e^* - m \chi_e^*) \right]  \cr
   &\qquad{} + \sum_{a \in \Sigma_\epsilon} \left[
    {1 \over 2} \psi_a (-i D^T \chi_a^* + M \psi_a^*) + 
    {1 \over 2} \psi_a^* (-i D^\dagger \chi_a - M \psi_a) \right]  \cr
   &\qquad{} - \sum_{e \in \partial\Sigma_\epsilon}
     \left[ {i \over 2} \chi_e^* \psi_{s(e)} + 
      {i \over 2} \chi_e \psi_{s(e)}^* \right] \,.
 }
This is an off-shell result.  On shell, the first two lines vanish.  By equating minus the on-shell action with the generating functional of connected Green's functions, we can see by following the logic of \cite{Iqbal:2009fd} that a sensible definition of the cutoff Fourier space Green's function $\tilde{G}_\epsilon(k)$ is
 \eqn{GtildePsi}{
  \chi_e(k) = -i \tilde{G}_\epsilon(k) \psi_{s(e)}(k) \,,
 }
provided $\tilde{G}_\epsilon(k)$ turns out to be real.  Assuming $m \neq 0$, we can rewrite \eno{GtildePsi} as
 \eqn{GtildeAgain}{
  {1 \over m} D\psi_e(k) = \tilde{G}_\epsilon(k) \psi_{s(e)}(k) 
 }
Plugging in the fermionic bulk-to-boundary propagator (identical to \eno{KakAnsatz} with $\phi_a \to \psi_a$ and $Q$ given by \eno{Qtrivial}), we obtain in place of \eno{CalculateGTilde} and \eno{GRen} the results
 \eqn{GRenAgain}{
  m\tilde{G}_\epsilon(k) &= {f_{v_\epsilon-1} - f_{v_\epsilon} \over f_{v_\epsilon}}
   = -{1 \over \zeta(\Delta-1)} + 
    |k|^{2\Delta-1} |\epsilon|^{2\Delta-1} {Q p^\Delta \over \zeta(2\Delta-1)} +
     \dots  \cr
  \tilde{G}(k) &= 
    \lim_{\epsilon \to 0} {\tilde{G}_\epsilon(k) \over |\epsilon|^{2\Delta-1}}
   = {Q p^\Delta \over m \zeta(2\Delta-1)} |k|^{2\Delta-1}
   = {\sqrt{p} \sgn k \over m \zeta(2\Delta-1)} |k|^{2\Delta-1} \,,
 }
where as usual we dropped the $k$-independent term from $\tilde{G}_\epsilon(k)$ before taking the $\epsilon \to 0$ limit.  The presence of the factor of $m$ in the denominator of the final expression in \eno{GRenAgain} makes sense because with a coupling
 \eqn{SintPsi}{
  S_{\rm int} = \int_{\mathbb{Q}_p} dx \, \left[ \psi_\text{def}(x)^* {\cal O}_\psi(x) + 
    {\cal O}_\psi(x)^\dagger \psi_\text{def}(x) \right]
 }
between a renormalized version $\psi_\text{def}$ of the boundary limit of $\psi$ and a boundary operator ${\cal O}_\psi$, the scaling symmetry must act as $\psi_\text{def} \to \lambda \psi_\text{def}$ and ${\cal O}_\psi \to (\lambda^*)^{-1} {\cal O}_\psi$.  So the two-point function $\tilde{G}(k)$ of ${\cal O}_\psi$ with ${\cal O}_\psi^\dagger$ scales as $\tilde{G}(k) \to |\lambda|^{-2} \tilde{G}(k)$, which matches the scaling of $1/m$.

In short, we arrive at
 \eqn{GxPsi}{
  G(x) = {u p^{2\Delta} \over m\zeta(2\Delta-1)} {\sgn x \over |x|^{2\Delta}} \,.
 }
It seems at first surprising that the symmetry of $G(x)$ under $x \to -x$ is the same for bosons and fermions.  The reason is that this symmetry was accompanied by complex conjugation, which reverses the order of operator multiplication without introducing signs related to the statistics of the operators.

\section{Gauge field dynamics}
\label{GAUGE}

So far we have considered only non-dynamical $U(1)$ gauge fields.  Let's now consider how we might add a kinetic term.  On a general directed graph $G$, define a face $f$ to be any subgraph of $G$ with three vertices only, all of which are required to be connected, let's say by edges $e_1$, $e_2$, and $e_3$.  To each face we assign (arbitrarily) a direction, meaning a direction around which we think of the boundary of the face circulating.  If the direction of edge $e_i$ matches the direction assigned to $f$, then we set $s(f,e_i)=1$; otherwise we set $s(f,e_i)=-1$.\footnote{This is of course the next step after \eno{dDef} in constructing the incidence matrix of the clique graph; see \cite{Knill:2013zz}.}  Then, starting from a function $\theta_e$ defined on directed edges, we set
 \eqn{dthetaDef}{
  d\theta_f = \sum_{i=1}^3 s(f,e_i) \theta_{e_i} \,.
 }
Because each $\theta_e$ is essentially a line integral $\int A_\mu dx^\mu$ over the corresponding edge, the derivative $d\theta_f$ is essentially the integral $\int {1 \over 2} F_{\mu\nu} dx^\mu \wedge dx^\nu$ over the face.  So an obvious analog of the Maxwell action is
 \eqn{Maxwell}{
  S_\theta = \sum_f {1 \over 2} (d\theta_f)^2 \,.
 }
Given a function $F_f$ on faces, we can immediately read off from \eno{dthetaDef} the adjoint operator (equivalently, the transpose):
 \eqn{dFDef}{
  d^\dagger F_e = \sum_{\partial f \ni e} s(f,e) F_f \,,
 }
where the sum is over all the faces whose boundary includes $e$.  Evidently---in the absence of couplings to other fields---the equation of motion following from \eno{Maxwell} is
 \eqn{ddMax}{
  d^\dagger d \theta_e = 0 \,.
 }
Although \eno{Maxwell} is indeed the obvious free field action for a $U(1)$ gauge field on a directed graph $G$, it is not entirely satisfactory, since we feel that the phases $e^{i\theta_e}$ rather than the angles $\theta_e$ should be the fundamental variables.  To say it another way, if $d\theta_f \in 2\pi \mathbb{Z}$ for some face $f$, we would like to say that it is equivalent to having $d\theta_f = 0$, but that is not reflected by the action \eno{Maxwell}.

The treatment is altogether more natural if we first define a Wilson line.  The task is not much harder for gauge group $U(n)$.  Starting therefore with a matter field $\phi$ mapping vertices to $\mathbb{C}^n$, we can define 
 \eqn{DphiU}{
  D\phi_e = U_e \phi_{t(e)} - \phi_{s(e)}
 }
where $U_e \in U(n)$.  Now consider a directed path $\gamma$ in our directed graph, from one vertex, $s(\gamma)$, to another, $t(\gamma)$.  It is simpler to rule out back-tracking and self-crossing, but this is not really necessary.  For edges $e \in \gamma$, let $s(\gamma,e) = 1$ if the direction of $e$ matches the overall direction of $\gamma$, and $-1$ otherwise.  Now define
 \eqn{UgammaDef}{
  U_\gamma = \prod_{e \in \gamma} U_e^{s(\gamma,e)} \,,
 }
where the order of factors follows the direction of $\gamma$: That is, the first edge in $\gamma$ corresponds to the leftmost factor, and the last edge corresponds to the rightmost factor.  The operator $U_\gamma$ is a Wilson line for the path $\gamma$.

Let $\partial_a f$ be the directed path around a face $f$, starting and ending on a chosen vertex $a$.  Then $U_{\partial_a f}$ is well defined, and using it we can form the action
 \eqn{SYM}{
  S_U = -{1 \over 2} \sum_f \tr(U_{\partial_a f} + U_{\partial_a f}^{-1}) \,,
 }
in analogy to the Yang-Mills action.  The choice of vertex $a$ on the boundary of each face can be made arbitrarily due to the cyclicity of the trace.  To find the equation of motion, choose an edge $e$ and consider a face $f$ with $e \in \partial f$.  One can check that $U_e$ is the first factor in $U^{s(f,e)}_{\partial_{s(e)} f}$.  Therefore, upon a variation
 \eqn{UeVariation}{
  U_e \to (1+i\alpha H) U_e \,,
 }
where $\alpha$ is a small parameter and $H$ is Hermitian, we find also $U^{s(f,e)}_{\partial_{s(e)} f} \to (1+i\alpha H) U^{s(f,e)}_{\partial_{s(e)} f}$.  We define the variation
 \eqn{deltaSU}{
  \delta S_U = \left. {\partial S_U \over \partial\alpha} \right|_{\alpha=0}
    = -{i \over 2} \sum_{\partial f \ni e} 
         \tr(H U^{s(f,e)}_{\partial_{s(e)} f}) + \text{c.c.}\,,
 }
where the sum is over all faces whose boundary includes $e$.  In short, the lattice Yang-Mills equation reads
 \eqn{LatticeYM}{
  \Im \sum_{\partial f \ni e} \tr(H U^{s(f,e)}_{\partial_{s(e)} f}) = 0 
    \qquad\text{for all hermitian $H$}\,.
 }

Returning to the $U(1)$ case, we can simplify the equations of motion \eno{LatticeYM} to
 \eqn{UoneLYM}{
  \Im \sum_{\partial f \ni e} U^{s(f,e)}_{\partial f} = 0 \,,
 }
where we omit to specify the starting and ending point of $\partial f$ because in an abelian theory it doesn't matter.  We wish to consider a stronger condition, which we will refer to as the complexified equation of motion:
 \eqn{StrongerYM}{
  \sum_{\partial f \ni e} U^{s(f,e)}_{\partial f} = 0 \,.
 }
Imposing the condition \eno{StrongerYM} clearly implies the equations of motion, but the reverse is in general not true.  This is reminiscent of the situation in continuum Yang-Mills theory where self-duality of the field strength implies the equations of motion, but not vice versa.  The point of interest for us is that the gauge field configurations we have used throughout do satisfy \eno{StrongerYM}.  To verify this, we need the identity
 \eqn{JacobiSum}{
  \sum_{y \in \mathbb{F}_p} \left( x-y \over p \right) \left( y \over p \right) = 
   (-1)^{p+1 \over 2} \,,
 }
where $x \in \mathbb{F}_p^\times$ is arbitrary and $p$ is an odd prime.  This convolution identity is known as a Jacobi sum, and it can be proven starting from the Gauss sum \eno{GaussSum}.  In the following two paragraphs, we explain how \eno{StrongerYM} reduces to \eno{JacobiSum}, as well as giving some further indications of what \eno{StrongerYM} means physically.

For $p \equiv 1 \mod 4$, the claim \eno{StrongerYM} comes down to the claim that $U_{\partial f} = +1$ for half of the $p-1$ faces adjoining a given edge $e$, and $-1$ for the other half.  Before we check this claim, let's note that since all the $U_e$ are real, the lattice Yang-Mills equation \eno{UoneLYM} is satisfied trivially.  It is not even clear that we should be demanding \eno{UoneLYM} if we restrict the gauge fields to $U_e = \pm 1$, since \eno{UoneLYM} was derived on the assumption that we could make an infinitesimal change to the gauge fields.  However, when all $U_e = \pm 1$, one can derive \eno{StrongerYM} as the condition that the action should remain unaltered when just one of the $U_e$ flips its sign.  In any case, checking \eno{StrongerYM} is trivial when $e$ is a vertical edge, because it comes down to the observation that half of the horizontal edges from any given vertex have sign $+1$ while the other have sign $-1$.  When $e$ is a horizontal edge, say from $0$ to $x \in \mathbb{F}_p^\times$, then we may write down \eno{StrongerYM} explicitly as
 \eqn{SYMexplicit}{
  \left( x \over p \right) + \sum_{y \in \mathbb{F}_p^\times \atop y \neq x}
   \left( x \over p \right) \left( y-x \over p \right) \left( -y \over p \right) = 0 \,.
 }
The first term in \eno{SYMexplicit} comes from a face with two vertical edges plus the chosen horizontal edge $e$, while the other $p-2$ terms come from faces with all edges horizontal.  The sum \eno{SYMexplicit} obviously reduces to \eno{JacobiSum}.

For $p \equiv 3 \mod 4$, because all the $U_e = i$, one finds that $U_{\partial f}$ is pure imaginary for all faces.  So in this case, \eno{StrongerYM} is equivalent to \eno{UoneLYM}, and we can think of it as following from requiring that the action is stationary under infinitesimal variations.  We do {\it not} see a useful way to understand how \eno{StrongerYM} arises from flipping the orientation of one edge while preserving the property that $U_e = i$, since any configuration in which all the $U_{\partial f}$ are pure imaginary automatically has vanishing action.  Once again, checking \eno{StrongerYM} is trivial when $e$ is a vertical edge because $U^{s(f,e)}_{\partial f} = i$ when the horizontal edge in $f$ points toward $e$ and $-i$ when it points away.  When $e$ is a horizontal edge, say from $0$ to $x \in \mathbb{F}_p^\times$, then \eno{StrongerYM} becomes
 \eqn{SYMthree}{
  i \left( x \over p \right) - i \sum_{y \in \mathbb{F}_p^\times \atop y \neq x}
    \left( x \over p \right) \left( y-x \over p \right) \left( -y \over p \right) = 0 \,,
 }
and this again reduces to \eno{JacobiSum}.

\section{Other sign characters}
\label{OTHERS}

Let's now inquire how we might modify our constructions on $L(T_p)$ so as to get the other three sign characters listed in section~\ref{INTRODUCTION}.  A standard notation is to specify a character by an element $\tau \in \mathbb{Q}_p^\times$ which is not a square.  Then
 \eqn{TauSpec}{
  \sgn_\tau x = \left\{ \seqalign{\span\TL &\qquad \span\TT}{
   1 & if $x=a^2-\tau b^2$ for some $a,b \in \mathbb{Q}_p$  \cr
   -1 & otherwise\,.} \right.
 }
Clearly $\sgn_\tau = \sgn_{\tau'}$ if $\tau/\tau'$ is a square.  As a result, there are essentially only four choices for $\tau$: $1$, $\epsilon$, $p$, and $p\epsilon$, where $\epsilon$ is any element of $\mathbb{F}_p^\times$ with $(\epsilon|p) = -1$.  The characters $\sgn_\tau$ can be explicitly evaluated, as follows:
 \eqn{TauExplicit}{
  \begin{tabular}{c||c}
   $p \equiv 1 \mod 4$ & $p \equiv 3 \mod 4$  \\ \hline\hline
   $\sgn_1 x = 1$ & $\sgn_1 x = 1$  \\
   $\sgn_\epsilon x = (-1)^{v_x}$ & $\sgn_\epsilon x = (-1)^{v_x}$  \\
   $\sgn_p x = (x_0|p)$ & $\sgn_p x = (-1)^{v_x} (x_0|p)$  \\
   $\sgn_{\epsilon p} x = (-1)^{v_x} (x_0|p)$ & $\sgn_{\epsilon p} x = (x_0|p)$
  \end{tabular}
 }
where $x = p^{v_x} (x_0 + x_1 p + x_2 p^2 + \dots) \in \mathbb{Q}_p^\times$ as in \eno{xExpress}, where $x_0 \in \mathbb{F}_p^\times$.

\begin{figure}
\centering{
\includegraphics[height=24ex]{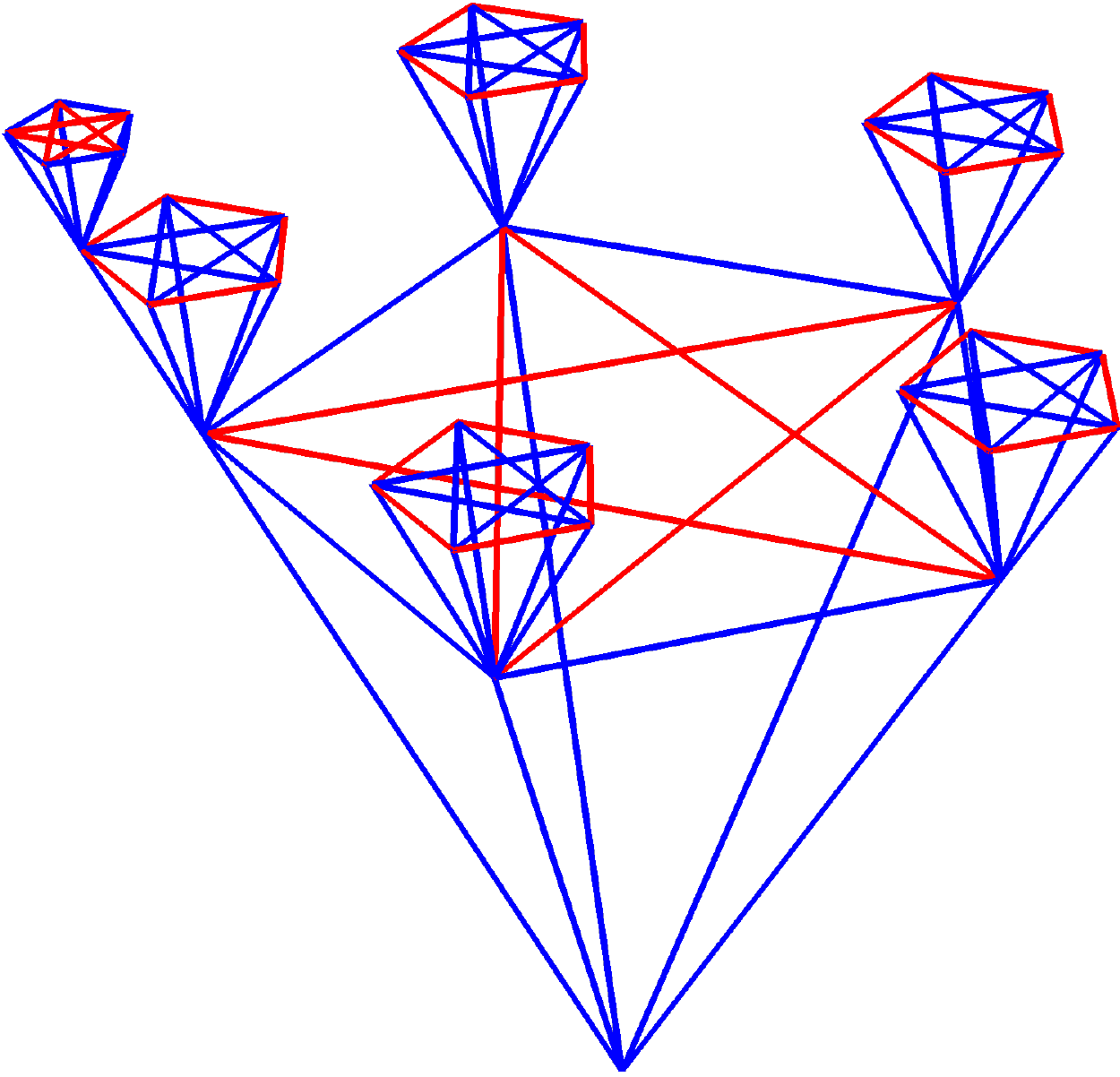} \hspace{3em}
\includegraphics[height=24ex]{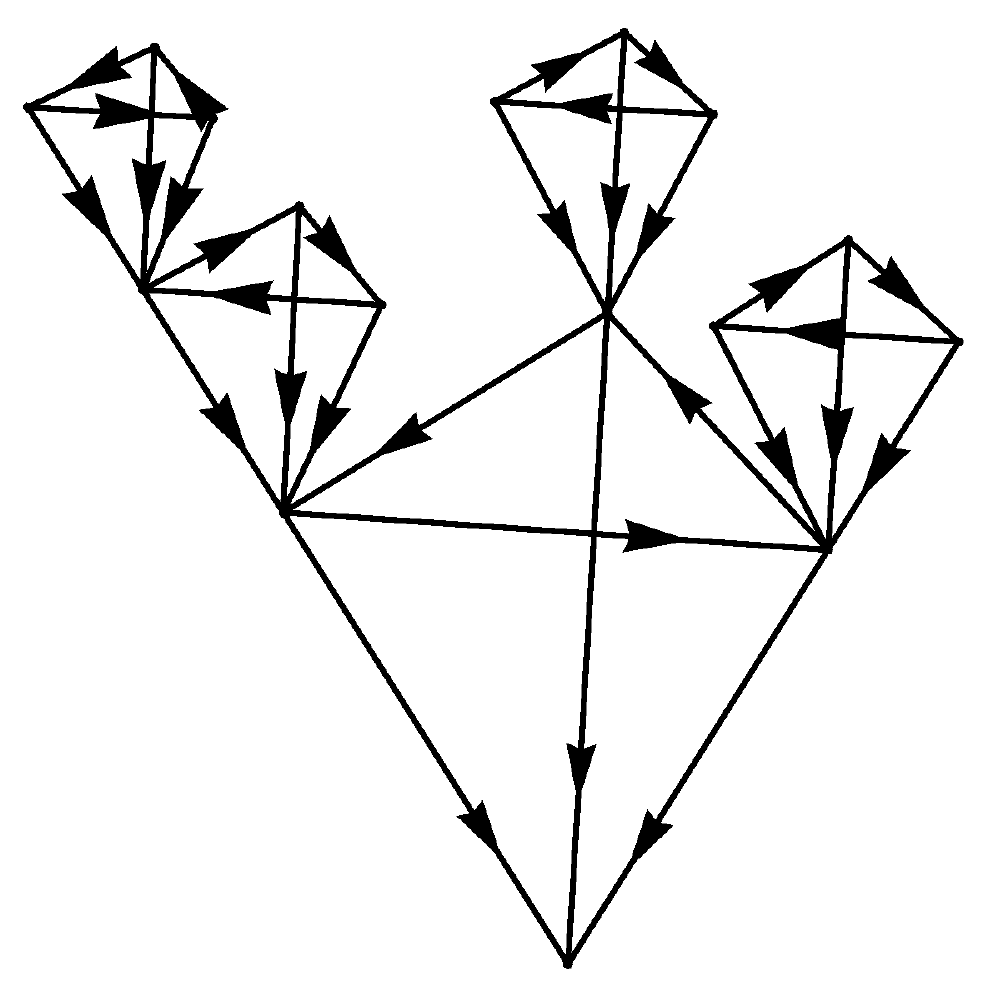} \hspace{3em}
\includegraphics[height=24ex]{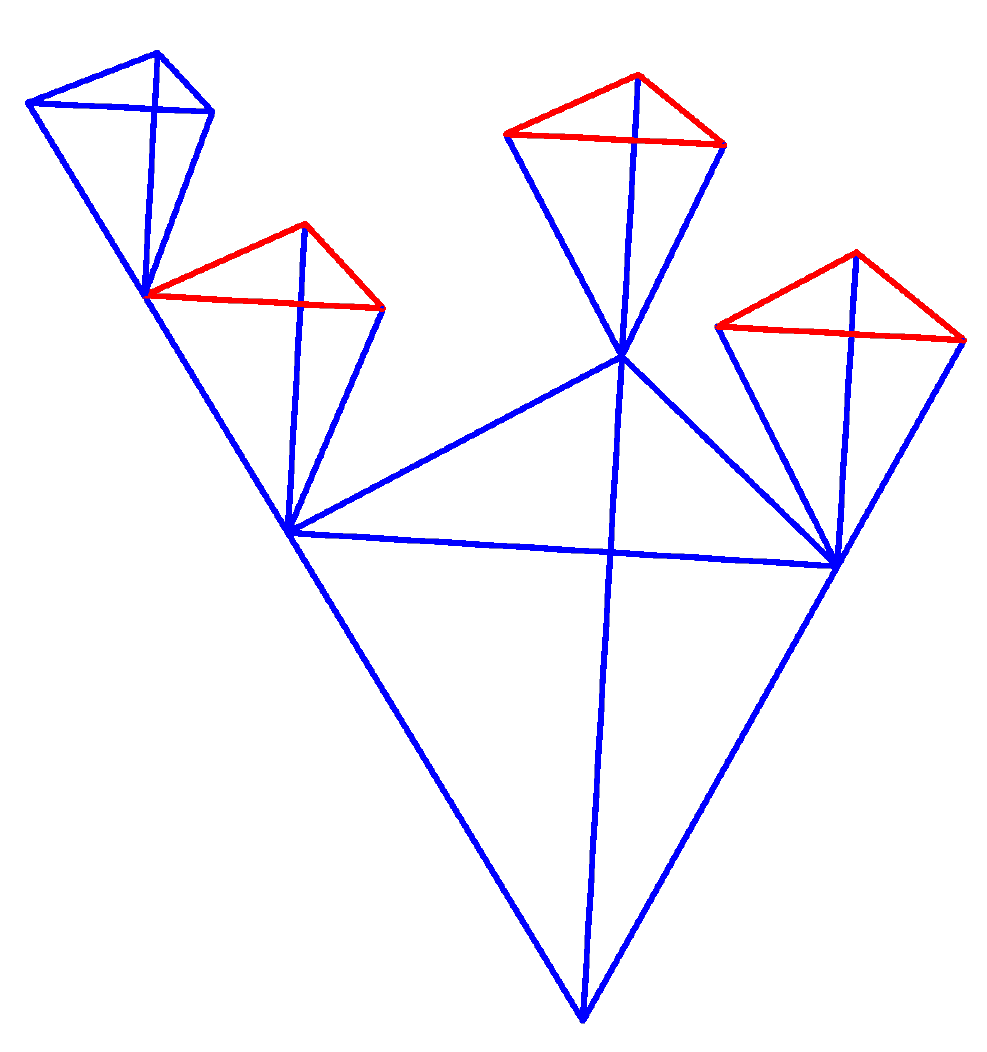}
}
\caption{
Left: $L(T_5)$ in blue and red.  The blue edges have $\theta_e = 0$, while the red edges have $\theta_e = \pi$. This geometry yields the sign character $(x_0|5)(-1)^{v_x}$.
\\  Center: $L(T_3)$ in black.  All the edges have $\theta_e = \pi/2$. This geometry yields the sign character $(x_0|3)(-1)^{v_x}$.
\\  Right: $L(T_3)$ in blue and red. The blue edges have $\theta_e = 0$, while the red edges have $\theta_e = \pi$. This geometry yields the sign character $(-1)^{v_x}$.} 
\label{LineGraphs2}
\end{figure}

A useful intuition in the way a sign character arises is that a particle picks up a phase $U_e$ as it propagates across a link $e$.  Because all vertical edges have the same downward orientation and the same $U_e$, any phase that a particle picks up while moving down from the boundary is precisely undone on its way back up.  However, on the shortest possible path on $L(T_p)$ from one boundary point to $\mathbb{Q}_p$ to another, there is one horizontal edge.  In a correlator $\langle O_\phi(x) O_\phi(0)^\dagger \rangle$, this horizontal edge runs from $(0,p^{v_x+1})$ to $(x_0 p^{v_x},p^{v_x+1})$.  So it makes sense that $\langle {\cal O}_\phi(x) {\cal O}_\phi(0)^\dagger \rangle$ should include a factor of
 \eqn{Udepth}{
  U_x \equiv \left\{ \seqalign{\span\TL &\qquad\span\TT}{
  U_{(0,p^{v_x+1}) \to (x_0 p^{v_x},p^{v_x+1})} & if 
   $(0,p^{v_x+1}) \to (x_0 p^{v_x},p^{v_x+1})$ exists as an edge  \cr
  U^{-1}_{(x_0 p^{v_x},p^{v_x+1}) \to (0,p^{v_x+1})} & otherwise\,.} \right.
 }
The factor \eno{Udepth} can be understood as a Wilson line for the path from $0$ to $x$---where all the factors for vertical edges canceled out.  The choice of non-dynamical $U(1)$ gauge fields in section~\ref{BACKGROUND} can be understood as a way to get $U_x = u \sgn x$---where the factor of $u$ is forced on us by the hermiticity condition $U_{-x} = U_x^*$.  It is not entirely clear from this discussion that the final Green's function $G(x)$ is proportional to $U_x$: For instance, other paths exist on $L(T_p)$ from $0$ to $x$ which do not pass through the horizontal edge in question.  However, it is a good guess that we can get the other sign characters in \eno{TauExplicit} by setting
 \eqn{HedgeU}{
  U_x = u_\tau \sgn_\tau x \,,
 }
where $u_\tau$ is an $x$-independent constant---either $1$ or $i$---chosen so as to preserve the condition $U_{-x} = U_x^*$.  If we require \eno{HedgeU} for all $x \in \mathbb{Q}_p$, and further require that phases should be invariant under translations, then \eno{HedgeU} amounts to a specification of $U_e$ for all horizontal edges.  See figure~\ref{LineGraphs2} for example.  As we have seen in the discussion following \eno{KakAnsatz}, we can gauge away the phase on vertical edges.

The calculations of previous sections are for the most part straightforward to generalize to arbitrary sign characters, using the prescription \eno{HedgeU} for choosing phases on the horizontal edges of $L(T_p)$.  For an operator dual to a complex scalar on $L(T_p)$ with action \eno{ScalarAction}, we find
 \eqn{GxGeneral}{
  G(x) = C {\sgn_\tau x \over |x|^{2\Delta}} \,,
 }
up to divergent contact terms proportional to $\delta(x)$.  We need $u_\tau = 1$ except when $p\equiv 3 \mod 4$ and $\tau = p$ or $\epsilon p$, and then $u_\tau = i$ is required.  The values of $C$ and $m^2$ can be determined from the following table:
 \eqn{KuDtable}{
  \begin{tabular}{c|c|c}
   $\tau$ & $C$ & mass-dimension relation  \\ \hline & & \\[-12pt]
   $1$ & ${p^\Delta \zeta(2\Delta) \over \zeta(2\Delta-1)^2}$ &
     $m^2 = -{1 \over \zeta(\Delta-1) \zeta(-\Delta)}$  \\[3pt]
   $\epsilon$ & $-{1+p^{2\Delta}-2p-m^2 \over 1+p^{2-2\Delta}-2p-m^2}
     {p^\Delta \zeta(4\Delta) \over \zeta(2\Delta) \zeta(4\Delta-2)}$ &
    $(2p+m^2)^2 = (1+p^{2\Delta})(1+p^{2-2\Delta})$  \\[3pt]
   $p$ & ${u p^{2\Delta} \over \zeta(2\Delta-1)}$ &
     $2p+m^2 = p^{1-\Delta} + p^\Delta$  \\[3pt]
   $\epsilon p$ & ${u p^{2\Delta} \over \zeta(2\Delta-1)}$ &
     $2p+m^2 = p^{1-\Delta} + p^\Delta$
  \end{tabular}
 }
where $u=1$ for $p \equiv 1 \mod 4$ and $u=i$ for $p\equiv 3 \mod 4$.  In all cases, we choose the larger of the two possible values of $\Delta$, assumed to be real.

The case $\tau=1$ is essentially the same as a complex scalar on $T_p$.  The cases $\tau=p$ and $\tau=\epsilon p$ are similar to one another, and we already presented in detail the cases $\tau=p$ for $p\equiv 1 \mod 4$ and $\tau=\epsilon p$ for $p\equiv 3 \mod 4$.

Let us therefore focus on the one case with some new features: $\tau=\epsilon$, i.e.~$\sgn_\epsilon x = (-1)^{v_x}$.  For simplicity, we consider only the complex scalar.  The treatment proceeds much as in sections~\ref{LINE} and~\ref{GREENLINE}, with a scalar ansatz
 \eqn{fAnsatz}{
  \phi_a = f_v \gamma(kz_a) \chi(-kx_a) \,.
 }
We recall the notation $z_a = p^v$, and we use $v_k$ as before to denote the valuation of $k$.  For $v > -v_k$, by plugging \eno{fAnsatz} into the scalar equation of motion \eno{EomPhi}, we obtain the difference equation
 \eqn{fDifference}{
  (2p+m^2) f_v - p f_{v+1} - f_{v-1} + (-1)^v (p-1) f_v = 0 \,.
 }
The last term comes from the horizontal edges.  The equation \eno{fDifference} is different from all previous difference equations we've encountered in that it does not have constant coefficients, but instead coefficients that are periodic modulo $2$ in $v$.\footnote{For the sign characters $(-1)^{v_x} (x_0|p)$, the difference equation obtained for $v>-v_k$ has constant coefficients because the terms from horizontal edges cancel out.}  Up to an overall multiplicative scaling, the general solution to \eno{fDifference} is
 \eqn{fSoln}{
  f_v = \left( 1 + q (-1)^v \right)
   \left( |z_a|^{1-\Delta} + Q |k|^{2\Delta-1} |z_a|^\Delta \right) \,,
 }
where $Q$ is a coefficient which at this stage is undetermined.  Plugging \eno{fSoln} into \eno{fDifference}, one finds
 \eqn{fmDeltaRelation}{
  (2p + m^2)^2 &= (1+p^{2\Delta})(1+p^{2-2\Delta})  \cr
  q &= {p^\Delta + p^{1-\Delta} - 2p - m^2 \over p-1} \,,
 }
and we assume as usual that $\Delta > 1/2$ is real.

For $v < -v_k$, the equation of motion \eno{EomPhi} is satisfied trivially, so the boundary condition that determines $Q$ comes from $v = -v_k$, where \eno{EomPhi} reads
 \eqn{fBC}{
  (2p+m^2) f_{-v_k} - p f_{-v_k+1} + 
   (-1)^{v_k} f_{-v_k} 
    \sum_{\alpha \in \mathbb{F}_p^\times} 
    \chi\left( -kz_a {\alpha \over p} \right) = 0 \,.
 }
Using the obvious identity
 \eqn{SimpleSum}{
  \sum_{\alpha \in \mathbb{F}_p^\times} 
   \chi\left( -kz_a {\alpha \over p} \right) = -1 \,,
 }
We arrive at
 \eqn{fBCExplicit}{
  \left[ 2p + m^2 - (-1)^{v_k} \right] f_{-v_k} - p f_{-v_k+1} = 0 \,.
 }
Plugging in \eno{fSoln} and using \eno{fmDeltaRelation}, we obtain
 \eqn{fQ}{
  Q = (-1)^{v_k+1} p^{1-2\Delta} {1+p^{2\Delta} - 2p - m^2 \over 
    1 + p^{2-2\Delta} - 2p - m^2} \,.
 }

To compute the holographic Green's function, if we start with \eno{CalculateGTilde}, we obtain
 \eqn{GTildeModulated}{
  \tilde{G}_\epsilon(k) = {1-q(-1)^{v_\epsilon} \over 1+q(-1)^{v_\epsilon}}
   \left[ p^{1-\Delta} + |k|^{2\Delta-1} |\epsilon|^{2\Delta-1} 
    {Qp^\Delta \over \zeta(2\Delta-1)} + \dots \right] - 1 \,, 
 }
where the omitted terms scale to $0$ more quickly than the ones shown as $\epsilon \to 0$ in the $p$-adic norm.  We now define
 \eqn{GRenDef}{
  \tilde{G}(k) = \lim_{\epsilon \to 0} \left(
    {1-q(-1)^{v_\epsilon} \over 1+q(-1)^{v_\epsilon}}
    {\tilde{G}_\epsilon(k) \over |\epsilon|^{2\Delta-1}} \right) = 
    {Qp^\Delta \over \zeta(2\Delta-1)} |k|^{2\Delta-1} \,,
 }
where we dropped a $k$-independent term from $\tilde{G}_\epsilon(k)$ before taking the limit.  As compared to \eno{GRen}, the definition \eno{GRenDef} may seem a bit contrived.  However, the extra prefactor ${1-q(-1)^{v_\epsilon} \over 1+q(-1)^{v_\epsilon}}$ in \eno{GRenDef} has no $k$-dependence, and its geometric mean between even and odd $v_\epsilon$ is $1$.  So we maintain that \eno{GRenDef} is the most sensible way to normalize the Green's function.  Passing through a Fourier transform, we wind up with
 \eqn{fGx}{
  G(x) = -{1+p^{2\Delta} - 2p - m^2 \over 1 + p^{2-2\Delta} - 2p - m^2}
    {p^\Delta \zeta(4\Delta) \over \zeta(2\Delta) \zeta(4\Delta-2)}
    {\sgn_\epsilon x \over |x|^{2\Delta}} \,.
 }

\section{Conclusions}
\label{CONCLUSIONS}

Our main results are summarized in \eno{HedgeU}-\eno{KuDtable}: With a suitably chosen configuration of a non-dynamical $U(1)$ gauge field on the line graph $L(T_p)$ of the Bruhat-Tits tree, we are able to recover from a bulk complex scalar action on $L(T_p)$ a holographic boundary two-point function proportional to any sign character one wants over $\mathbb{Q}_p$---for any odd prime $p$.  Equally, we can work with a bulk complex fermion.  The key is not the statistics of the gauge field (or the boundary operator), but rather the bulk gauge field, which gives rise to the desired sign character essentially as a Wilson line between two boundary points.

While technical in nature, our results open up many further questions.  To begin with, we would like to work out the case $p=2$.  There are seven non-trivial sign characters $\sgn_\tau x$ over $\mathbb{Q}_2$ (besides the trivial one), and they depend not only on the $2$-adic norm of $x$, but also on its second and third non-trivial $2$-adic digits.  Preliminary indications are that we can engineer elaborations of $T_2$, including non-dynamical gauge fields, that allow us to recover these sign characters; however, nearest neighbor interactions are not enough.  This is not too surprising given that the second and third $2$-adic digits relate to paths on the tree with at least two or three links.  Perhaps this is a hint for how to go on to more complicated multiplicative characters over $\mathbb{Q}_p$, which can depend on finitely many $p$-adic digits and will wind up involving finite range interactions on $T_p$.

It is a bit unsatisfying that we have placed so few limitations on the types of fields that are allowed on $T_p$.  In particular, if on $L(T_p)$ we give each horizontal edge a phase $\theta_e = \pi$ and leave all vertical edges with phases $\theta_e = 0$, then we wind up with a correlator of the form \eno{GxGeneral} with the trivial sign character, and the only effect of the phases is to switch the sign of $C$.  We would like to think that there is a positivity constraint that fixes this sign, but we do not know how to articulate it.  We could observe that the choice of phases just described fails to satisfied the complexified equation of motion \eno{StrongerYM}; however, the same critique can be made of the choice we made to capture the character $\sgn_\epsilon x = (-1)^{v_x}$.  A more fundamental point of view is called for, explaining why particular choices of the $U(1)$ gauge field are natural constructions on $L(T_p)$, perhaps analogous to the way that the spin connection is natural on a smooth manifold.

Perhaps a related point is that we seem to have broken a lot of the symmetry of $T_p$ by introducing the distinction between horizontal and vertical edges on $L(T_p)$.  At one level, this is not so disturbing, because the invariance we should require is dual to the maps $x \to rx + b$ for all $b \in \mathbb{Q}_p$ and $r \in (\mathbb{Q}_p^\times)^2$.  From a boundary theory point of view, the operators ${\cal O}_\phi$ and their two-point Green's function transform with non-trivial Jacobians under other elements of the $p$-adic conformal group ${\rm PGL}(2,\mathbb{Q}_p)$.  It would be very satisfying to give a full account in the bulk of how the corresponding isometries of $T_p$ (and $L(T_p)$) act on the $U(1)$ gauge fields so as produce holographic Green's functions which are suitably covariant under the $p$-adic conformal group.

Having introduced the possibility of gauge field dynamics on $L(T_p)$, another natural direction to explore is what the corresponding boundary operators are.  If some notion of conserved currents on the boundary is understood, perhaps the gravitational dynamics of \cite{Gubser:2016htz} could be refined or extended.  We also hope that an enriched understanding of the geometry dual to $p$-adic conformal field theories will eventually impact back on $p$-adic string theory, perhaps providing a better first-principles understanding of Freund and Olson's adaptation of the Veneziano amplitude and suggesting some interesting generalizations.

\section*{Acknowledgments}

We thank M.~Heydeman for substantial input in early phases of this work, in particular for identifying the $U(1)$ gauge field configurations as related to Paley graphs.  This work was supported in part by the Department of Energy under Grant No.~DE-FG02-91ER40671, and by the Simons Foundation, Grant 511167 (SSG).

\bibliographystyle{utphys}
\bibliography{fermion}

\end{document}